\documentclass[aps,prb,reprint]{revtex4-2}
\usepackage[hypertexnames=false]{hyperref}
\usepackage{amsmath,bm}
\usepackage{xcolor}
\usepackage{graphicx}
\graphicspath{{BaCoSiO4_draft_-_Copia}}
\begin{document}
\title{Interplay of trimerization, chirality and ferroelecticity in multiferroic BaCoSiO$_4$}

\date{January 2022}

\author{Aditya Putatunda}
\author{Sergey Artyukhin}
\affiliation{Istituto Italiano di Tecnologia, Genova 16163, Italy}
\begin{abstract}
    Multiferroic materials combine multiple ferroic orders that may enable cross-functionalities, e.g. control of one ferroic order by the field conjugate to the other one. BaCoSiO$_4$, a recently proposed multiferroic combines multiple unit cell-tripling modes, structural chirality and ferroelectric polarization, whose interactions result in a peculiar ground state. We derive the Landau free energy for interacting trimerization, electric polarization and structural chiral modes and find the corresponding coefficients from first-principles calculations. This results in a quantitative model for the description of domain walls and vortices, observed in recent experiments.
\end{abstract}
\maketitle
\section{Introduction}
Multiferroics host multiple orders, whose interactions provide a playground for switching one order with a field, conjugate to another, as well as for their peculiar domain structures \cite{Spaldin2019advances}. A particular place is reserved for improper ferroelectrics, where ferroelectricity is induced indirectly by another order, thus ensuring strong coupling between them, as in spiral multiferroics \cite{Cheong2007multiferroics}. Improper ferroelectrics have peculiar domain structures, with domain walls in multiple order parameters clamped together~\cite{Hanamura2003clamping,Giraldo2021magnetoelectric,Artyukhin2014landau}. In this context, hexagonal manganites and ferrites have unit cell tripling structural distortions, that break inversion and induce ferroelectric polarization \cite{Choi2010insulating,Artyukhin2014landau}. The free energy having the Mexican hat shape with multiple domains along the rim gives rise to fascinating networks of domain walls and vortices that resemble cosmic strings in their critical dynamics \cite{Griffin2012}.  
In this paper, we discuss the fascinating interplay of ferroic orders in BaCoSiO$_4$, recently reported as a qualitatively new multiferroic demonstrating such a structural trimerization~\cite{Ding2021field}. It constitutes a particularly interesting example comprised of unusual coupling between magnetism, chirality, polarization and structural trimerization~\cite{Ding2021field,Xu2022multiple,Cheong2022magnetic}. Interplay of multiple orders in such materials have inspired the development of symmetry-operational similarity (SOS) approach, that helps to rationalize their resulting complex phenomena.\cite{cheong2019sos,cheong2021permutable}

\begin{figure}[b]
    \centering
    \includegraphics[width=\linewidth]{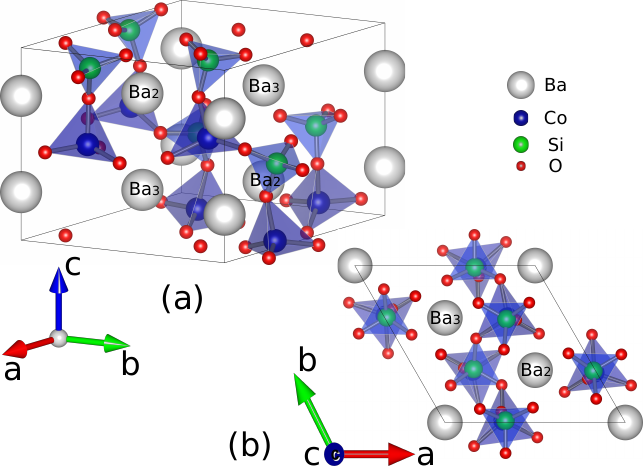}
    \caption{Crystal structure of BaCoSiO$_4$ at 300 K showing its trimerized P6$_3$ symmetry as viewed from the (111) and (001) directions (panels (a) and  (b), respectively). The green, blue, grey and red spheres respectively represent the Si, Co, Ba and O atoms. The tilting of corner-sharing tetrahedra surrounding Ba ions drive their displacements along the $c$-axis to accommodate the tilts. The slight off-centering shifts of equatorial oxygens in a clockwise circular motion for three neighboring tetrahedral units within the $ab$-plane are seen in panel (b).}
    \label{fig:struct}
\end{figure}

BaCoSiO$_4$ is experimentally observed to crystallize in a stuffed tridymite structure \cite{Liu1993structures} with its symmetry described by the space group P6$_3$ (\#173 of International Tables) as shown in Fig.~\ref{fig:struct}. The Co$^{2+}$ (bigger) and Si$^{4+}$ (smaller) ions are positioned inside corner-sharing, thus resulting into tilted tetrahedra of different sizes. The Ba$^{2+}$ ions stuff the equatorial voids between these tetrahedra, shifting slightly along the $c$-axis to accommodate their tilt.
Around 530$^\circ$~C the cell volume tripling distortion (trimerization) sets in. With it, three neighboring systems of corner-sharing SiO$_4^{4-}$ and CoO$_4^{6-}$ tetrahedra tilt, thus bending the Si-O-Co bonds by about 32$^{\circ}$ away from the ideal straight shape, as observed experimentally.\cite{Liu1993structures}. The apical oxygens, corner-shared between the Si and Co tetrahedra, shift in the $ab$ plane, with the shifts of the neighboring oxygens at 120$^\circ$ to each other, as indicated in panel (b), Fig.~\ref{fig:struct}. The structural transition may be analogous to octahedral tilting transitions in ABO$_4$ perovskites that occur due to size mismatch between the large $A$-site ions and $B$-centered octahedra. In an undistorted BaCoSiO$_4$, there are columns made up by two large Ba ions within the unit cell in the $c$ crystallographic direction, with 4 ionic radii $r_A$ making up for a translation vector. On the other hand, four corner-shared tetrahedra also span the unit cell vertically, so that the $c$ lattice constant must match with 4 tetrahedron heights $h_T=\frac{4}{3}(r_B+r_X)$, with $r_B$ being the radii of Si and Co cations, while $r_X$ -- of oxygen anions. The tolerance factor $t=r_A/h_T=\frac{3r_A}{4(r_B+r_X)}=$ 1 indicates an ideal match among their ionic radii, favoring no trimerization. For $t<1$, the $c$ lattice constant measured through the network of tetrahedra is larger than that through Ba columns, hence buckling of the tetrahedra helps equalize the two. Using Shannon appropriate radii values for each species, we obtain $t\approx$ 0.95.\cite{shannon1976}

Earlier, Taniguchi and co-workers reported a 300\% enhanced dielectric permittivity below 1~kHz in BaCoSiO$_4$ under photoexcitation at 365 nm. It is attributed to photoexcited electrons trapped in the localized conduction Co-$d$-orbitals.\cite{Taniguchi2014} The alternating Co and Si occupations and rotations of tetrahedra make this material both chiral and polar. S.~Yu suggested that its chiral structure combined with the magnetic order in Co$^{2+}$ ions causes non-reciprocal propagation of light showing magneto-chiral dichroism.\cite{Yu2019}, as evident in Fig.\ref{fig:walls}, panel (b) for its propagation vector along the $c$-axis. In their most recent investigation, Ding and co-workers~\cite{Ding2021field} reported the  observation of a vortex-like spin texture, where using magnetic field, they were able to tune its net toroidal moment along a series of metamagnetic transitions. It has been proposed that the toroidal moment, arising frustrated antiferromagnetic interactions of the spin magnetic moments arranged at 120$^{\circ}$ spin vortex on three interpenetrating magnetic sublattices of Co atoms form the material's magnetic ground state. Such a transition, taking account of the both broken inversion and time-reversal symmetries in BaCoSiO$_4$, as discussed in this paper, is bound to give rise to electric polarization as reported by the magnetoelectricity observed Xu and co-workers.\cite{Xu2022multiple,Xu2023large}

The paper is structured as follows: we start by systematically reviewing structural orders starting from a high-symmetry reference parent structure found using pseudosymmetry search. We discuss the symmetry lowering by the present distortions, and formulate a Landau-type theory that describes their interactions. Finally we review possible domains and domain walls.

\section{Symmetry analysis}
In this study, we aim to understand the relationship between different structural orders in BaCoSiO$_4$ and their possible associated functionalities. In order to classify the lattice distortions, we proceed by finding the parent high-symmetry structure. Using Pseudosymmetry application \cite{Capillas2011} from Bilbao crystallographic server \cite{Aroyo2006bilbao1,Aroyo2006bilbao2,Aroyo2011crystallography}, we identified the P$6_3/mmc$ (\#194 of International Tables) parent structure  (refer Fig.~\ref{fig:symmetry}), where occupations of sites inside the tetrahedra are uniformly disordered between Si and Co. With respect to this reference, the experimental room-temperature structure has the following structural orders: structural chirality $\mathcal{C}$, polarization $P_z$, trimerization $K_1 \ldots K_4$. By using the centrosymmetric and non-chiral reference structure, we can calssify the trimerizing distortions into odd and even components with respect to inversion and mirror symmetries, thus obtaining four two-component order parameters, $K_1\ldots K_4$.

\begin{figure}[b]
    \centering
    \includegraphics[width=.6\linewidth]{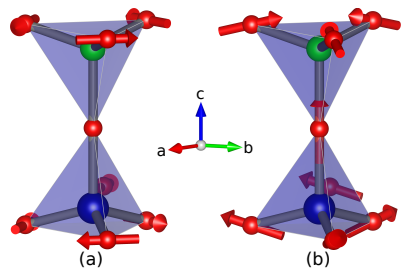}
    \caption{Distortions on the Si and Co-centered tetrahedra, corresponding to two zone center modes, $\Gamma_1^-$, shown in panel (a), and $\Gamma_2^-$, in panel (b). Arrows indicate the ionic displacements.}
    \label{fig:gamma_modes}
\end{figure}

The high symmetry parent structure, starting from the right end, panel (d) of Fig.~\ref{fig:symmetry} describes a possible chain of resulting intermediate structures. As one moves from panel (d)-(a), its global symmetries are gradually lowered  through symmetry adapted displacements as represented by the indicated irreducible representations on the connecting arrows. Acting with the chiral $\Gamma_1^-$ mode, causes the individual Co/Si tetrahedra to counter-rotate about their $c$ axis to yield a P6$_3$22 [panel (c)] (\#182) structure. Further displacing its ions according to the chiral $\Gamma_2^-$ mode results into restoring the ionic order as present in the experimental structure, making the Co-O-Si bond polar in untrimerized P6$_3$ structure, panel (b). Finally, the $K_1$ mode distortion causes trimerization with tripling of the unit-cell volume, thus yielding the experimentally observed ground state. It must be noted that Fig.~\ref{fig:symmetry} represents only one of the possible structure chains that yields physically relevant intermediate structures. One may also obtain other structures using a different combination order of these displacement modes but it might yield highly distorted, hence physically unrealizable structures.

\begin{figure*}[t]
    \centering
    \includegraphics[width=\linewidth]{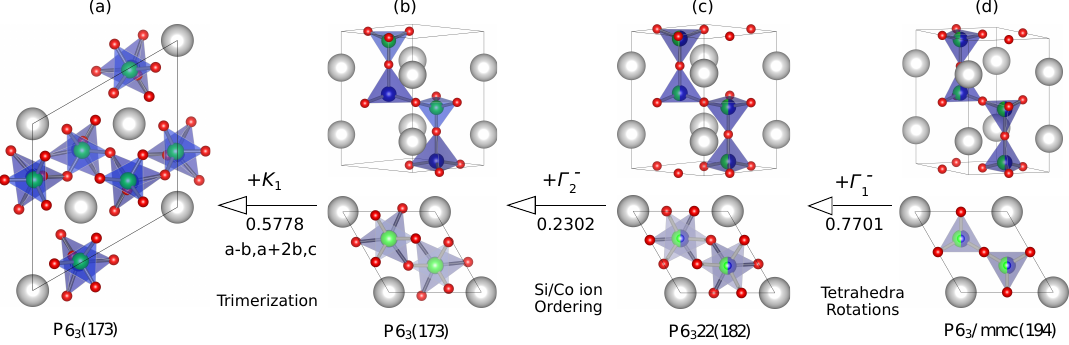}
    \caption{Group-subgroup relationships and a possible pseudosymmetry structures for BaCoSiO$_4$. Low-temperature trimerized P$6_3$ structure is shown one far left [panel (a)].}
    \label{fig:symmetry}
\end{figure*}

\subsection{Identifying the distortion modes}
Group theoretical analysis using ISODISTORT \cite{Stokes2006isodisplace, stokes2019isodistort} between the distorted structure and its symmetric parent yields six distortion modes, namely, the two zone centre modes: $\Gamma_1^-$, $\Gamma_2^-$ and four zone-boundary modes, $K_1\ldots K_4$. In the following we inspect each of the modes, illustrated in detail in Fig.~\ref{fig:k_combo}.

{\it Zone center modes --- }
\textbf{$\Gamma_1^-$} describes structural chirality and involves the counterrotations of CoO$_4$ and SiO$_4$ tetrahedra, as shown in FIG.~\ref{fig:gamma_modes}(a). Namely, equatorial oxygen atoms in CoO$_4$ and SiO$_4$ rotate within the $ab$ plane around Co-Si axes clockwise and counterclockwise, respectively.

Polar \textbf{$\Gamma_2^-$} mode corresponds to Co/Si ionic ordering, associated volume distortions of the tetrahedra, and apical oxygen off-centering along the bond. Size disparity between Co and Si ions makes Co and Si-centered tetrahedra bigger and smaller, respectively, thus driving the shift of apical oxygen atoms. This mode is also accompanied by much smaller off-centering for all Co and Si atom pairs as well as Ba ions, which are omitted for simplicity in FIG.~\ref{fig:gamma_modes}(b).

{\it Zone boundary modes (trimerization) ---}
With respect to a centrosymmetric parent structure as a reference, the components of trimerizing mode split into inversion-odd and even and mirror odd and even, thus resulting in four different modes. Here we review them all. In $K_1$ mode, the largest displacements involve the apical oxygen atoms, which are followed by smaller shifts to the equatorial ones all within the $ab$ plane. As shown in FIG.~\ref{fig:k1_k2}(a), the apical oxygen atoms, in groups of three, move closer to the Ba atoms on the cell boundary whereas they move away from the four Ba atoms in the center of the cell. This symmetry breaking can be viewed as trimerization, however, there exists no movement corresponding to off-centering of the Ba atoms themselves in this mode.

K$_2$ mode overall has the smallest magnitude of distortion, involves displacements of the equatorial oxygens within the $ab$ plane only, and includes a smaller but opposite off-centering of the Si and Co cations included with an off-centering of only the cell-centered Ba atoms along the $c$-direction.
    
The $MO_4$ tetrahedra distorts in a way such that one oxygen out of the three equatorial ones cooperatively moves closer corresponding to void created by the Ba off-centering. The Co/Si atoms from the three neighboring tetrahedra move closer towards these Ba atoms, not shown for their smaller magnitude. The details have been plotted in FIG.~\ref{fig:k_combo}(b).
    
K$_3$ mode excludes the displacements of the apical oxygens. The outer Ba atoms displace twice as much as the central ones but in the opposite directions along the $c$ direction, thus keeping their overall center of mass fixed. Each pair of three Co or Si atoms for a given $c$ plane as well as for two adjacent planes off-center in an opposite circular sense about their central Ba atoms as shown in FIG.~\ref{fig:k_combo}(c). The asymmetric equatorial oxygen displacement, with a small $c$ component, follows this direction of Co/Si displacement.
    
K$_4$: As indicated by the ISOVIZ\cite{stokes2019isodistort} tool, this mode has the largest distortions. For the apical oxygens, the main displacements are in the $ab$ plane and orthogonal to those in the $K_1$ mode, as shown in FIG.~\ref{fig:k_combo}(d). For comparison of the displacements in the $K_1$ and $K_4$ modes, refer to FIG.~\ref{fig:k1_k2}(b). While the $K_1$ mode displaces the apical oxygens towards the central Ba atoms, $K_4$ essentially involves moving them in a circular fashion around the same center.
    
Also, two of the three equatorial oxygen atoms in a given tetrahedron move in a similar direction, keeping the third one fixed. This leads to a tilting of these rigid units apart within a trimer.
    
\section{Landau theory}
\begin{figure*}[t]
    \centering
    \includegraphics[width=\linewidth]{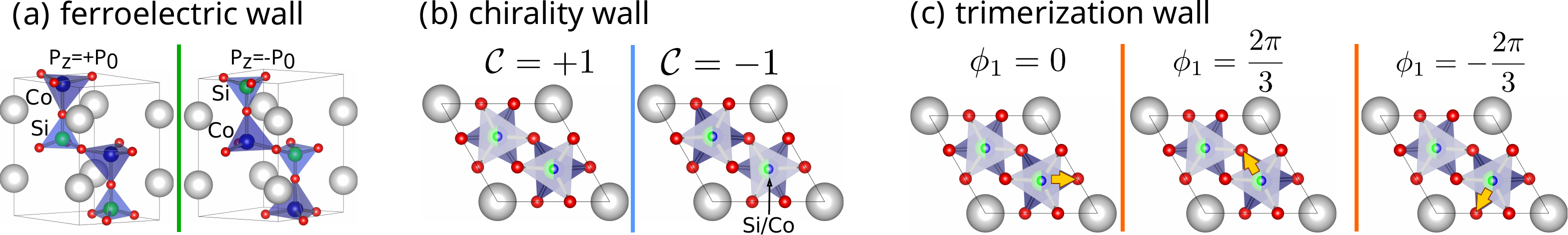}
    \caption{Structural domain walls, realized in BaCoSiO$_4$. (a) Ferroelectric domain wall, separating domains with two possible of Si-Co ionic ordering. (b) Domain wall between two chiral domains with distinct pyramidal rotations. (c) trimerization domain wall, with trimerizing displacements of corresponding ions in neighboring domains pointing at 120$^\circ$.}
    \label{fig:walls}
\end{figure*}

We introduce one-dimensional order parameters $\mathcal{C}, P_z$ transforming according to $\Gamma_{1,2}^-$, respectively, and representing the chirality and the $z$ component of the ferroelectric polarization: P$_z$, and two-dimensional order parameters $\vec Q_{1\dots 4}$ transforming as $K_1\dots K_4$ and representing trimerization. The domain walls, corresponding to the reversal of these order parameters, are shown in Fig.~\ref{fig:walls}.

Since the anisotropy in hexagonal materials starts from six-order terms, and therefore the trimerization amplitude does not change significantly across domain walls \cite{Artyukhin2014landau},  it is convenient to use polar coordinates ${\vec Q_n=Q_n(\cos\phi_n,\sin\phi_n)}$.
Treating $\mathcal{C}$, $P_z$ and $\vec Q_1$ as primary order parameters, we proceed to analyze the possible domains and domain walls.
Second order part of the free energy thus takes the form,
\begin{equation}
\mathcal{F}_2=\frac{1}{2}\left(\alpha_1 Q_1^2+\alpha_2 Q_2^2+\alpha_3 Q_3^2+\alpha_4 Q_4^2\right),\label{F2}
\end{equation}
where $\alpha_1<0$ and $\alpha_{2,3,4}>0$.
Degree 3 invariant polynomials are found and the free energy expressed in their terms as:
\begin{multline}
\mathcal F_3=\beta_1Q_1^3\cos 3\phi_1+
g_2\mathcal{C}\vec Q_1\cdot \vec Q_2+
g_{12}Q_1Q_2^2\cos(\phi_1+2\phi_2)\\
+g_3P_z\vec Q_1\cdot \vec Q_3+
g_{13}Q_1Q_3^2\cos(\phi_1+2\phi_3)\label{F3}
\end{multline}

The term with $\beta_1$ describes the anisotropy for $\vec Q_1$, and results in three minima at $\phi_1=0,\pm \frac{2\pi}{3}$ for $\beta_1<1$ and in the minima at $\phi_1=\pi$ and $\pm \frac{\pi}{3}$ otherwise. Since the mode decomposition of the experimental structure (using ISODISTORT tool \cite{Stokes2006isodisplace,stokes2019isodistort}) gives $\phi_1=0$, $\beta_1$ must be negative.
The term with $g_2$ determines the slave order $\vec Q_2$. Indeed, minimizing Eq.~\ref{F2}-\ref{F3} with respect to it, we obtain $\vec Q_2=-\frac{g_2}{\alpha_2}\mathcal{C}\vec Q_1$.
Similarly, from minimizing the $g_3$ term along with Eq.~\ref{F2}, we have $\vec Q_3=-\frac{g_3}{\alpha_3}P_z\vec Q_1$.
The interaction between $P_z$, $Q_1$ and $Q_3$ implies that in a crystal where $P_z$ domain wall is created during the growth, $\vec Q_1\cdot \vec Q_3$ must change across that wall, i.e. the ferroelectric domain wall must coincide with the trimerization wall.

The interactions, responsible for setting the direction of $\vec Q_4$ are contained in the fifth order terms:
\begin{equation}
    \mathcal F_4=g_4\mathcal{C} P_z Q_1^2Q_4\cos(2\phi_1+\phi_4)+\beta_4\mathcal{C} P_zQ_4^3\cos (3\phi_4).\label{FQ4}
\end{equation}
Minimizing this together with Eq.~\ref{F2}, we obtain 
\begin{equation}
    \vec Q_4=-\frac{g_4}{\alpha_4}\mathcal{C} P_z Q_1^2\binom{\cos 2\phi_1}{-\sin 2\phi_1}.
\end{equation}
This peculiar dependence on $2\phi_1$ implies that the slave order parameter $\vec Q_4$ turns twice faster than $\vec Q_1$ across domain walls.

\vskip 10pt
{\it Induced strain ---} The isotropy subgroup for an arbitrary direction of $\vec{Q}_1$ is P3 (\#143) which retains the 3-fold symmetry of the experimental structure P6$_3$ (\#173). However, shear strain within the $ab$-plane, $\epsilon_{xy}$ and $\epsilon_{xx}-\epsilon_{yy}$, as well as $\epsilon_{xz}$ and $\epsilon_{yz}$ would break the threefold axis, and hence are not allowed. In fact, the displacements of apical oxygens do break the C$_3$ axis of the corner-sharing tetrahedra, but having two other units with 120$^\circ$ rotated shifts preserves C$_3$ symmetry by mapping them onto each other. Any shear strain, for example, $\epsilon_{xy}$, would then break this three fold symmetry. This implies that the shear strain cannot be induced by the $\vec{Q}_1$ mode.

{\it Antiferromagnetic ordering ---} Co$^{2+}$ ions adopt a 120$^\circ$ antiferromagnetic ordering below $T_N$=~3.2~K. Trimerization then sets a preferred easy axis direction for spins, which is communicated to the spin sector via single-ion and DM interactions. The magnetic point group is 173.129 \cite{Ding2021field}, corresponding to the active magnetic irreducible representation mK$_1$ in the direction $(a,0)$ frozen into the paramagnetic parent P$6_3$ (\#173). Following \cite{Ding2021field}, we denote the corresponding order parameters, toroidal moments for the neighboring layers, as $\bm{T}_1, \bm{T}_2, \bm{T}_3$. The term in the free energy: 
\begin{equation}
    \mathcal{F}_{\mathrm{T}}=g_TP_zH_z(T_{1z}+T_{2z}+T_{3z})+\sum_i\kappa(\bm{T}_i\cdot\bm{Q})^2
\end{equation}
results in the weak ferromagnetic moment $M_z=-\partial \mathcal{F}/\partial H_z=-g_tP_z(T_1+T_2+T_3)$, whose direction is determined by the sign of the ferroelectric polarization $P_z$ and the toroidal moments. A progressively increased external magnetic field $H_z$ favors states with larger $(T_1+T_2+T_3)$, thus leading to a sequence of transitions between the states with increasing $M_z$ \cite{Ding2021field}. The term with $\kappa$ ties the magnetic easy axis to the direction of the trimerization. Thus, at a structural domain wall, where the trimerization rotates by 120$^\circ$, the antiferromagnetic order parameter has to rotate by $+120^\circ$ or $-240^\circ$ for the spins across the wall to align with the easy axis.

\section{Domain walls and vortices}
\begin{figure*}[t]
    \centering
    \includegraphics[keepaspectratio,angle=-90,width=0.98\textwidth]{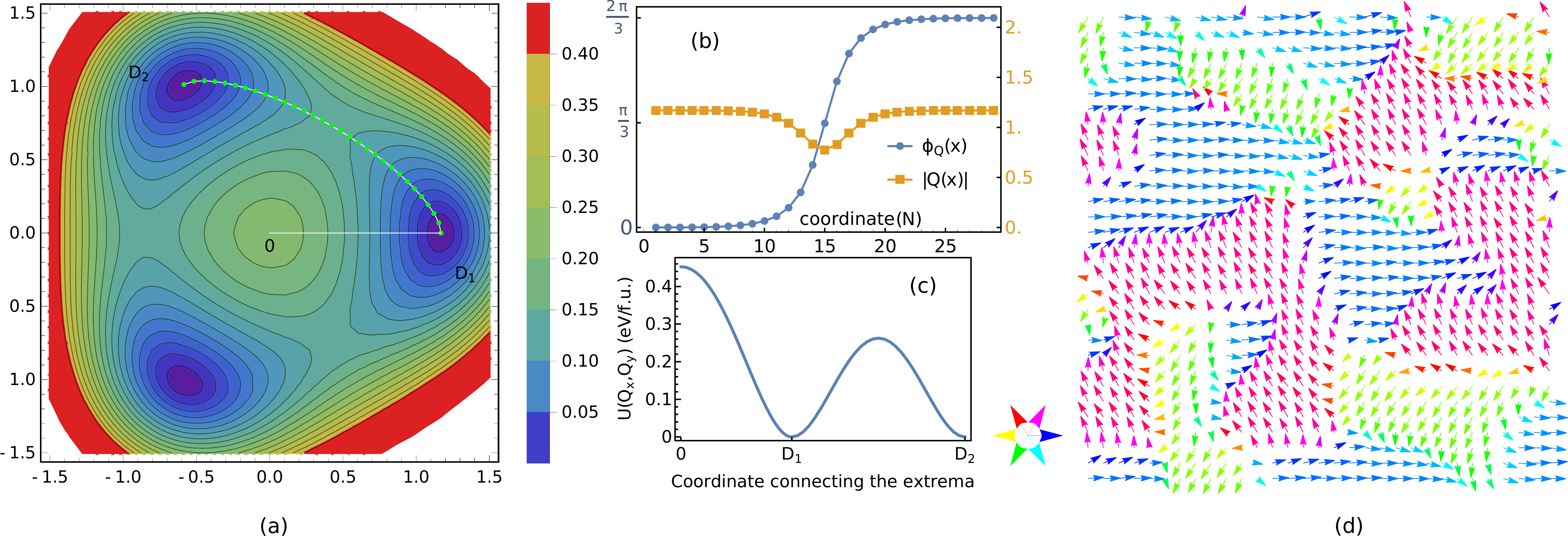}
    \caption{(a) Potential energy landscape as a function of trimerization components $F(Q_x,Q_y)$ and its cuts (c) along the path $0-D_1-D_2$, indicated in panel (a). (d) The spatial profiles of trimerization amplitude and angle across a trimerization domain wall. A snapshot of a phase field simulation, showing the formation of domain walls and triangular vortices. The length of the arrows represent the magnitude of trimerization whereas the color codes can be read from the legend such that blue represents trimerization at 0$^\circ$}
    \label{fig:barrier-specs}
\end{figure*}

In this section we study the structure of the possible domain walls. By combining the Landau theory and ab-initio simulations, we aim to obtain a realistic estimate for the wall structure and energy. Since polarization and chiral modes correspond to large displacements, such walls will likely be sharp and immovable, and the trimerization walls and vortices, discovered in recent experiments, may be the most relevant. For the purposes of modelling the trimerization wall in the volume of the material where $P_z$ and $C$ are constant, we here write the free energy due to the order parameter: $Q$ = ($Q_x$, $Q_y$) that connects untrimerized reference structure P6$_3$ [with $P_z$ and $C$ present, Fig.~\ref{fig:symmetry}(b)] to a trimerized low-symmetry P6$_3$, illustrated in Fig.~\ref{fig:symmetry}(a). In this group-subgroup hierarchy, all the $K_1\ldots K_4$ correspond to a single 2-dimensional representation $K_1$ in P$6_3$ which we denote as: $Q=(Q_x,Q_y)=Q(\cos\phi,\sin\phi)$. The free energy density has a form of an upward facing Mexican hat potential with 3-fold symmetry (Fig.~\ref{fig:barrier-specs}),
\begin{equation}
    \mathcal{F} = c(\nabla\bm{Q})^2-aQ^2+bQ^4+\gamma Q^3\cos3\phi.
\end{equation}

Here, the first two terms give a rotationally-symmetric Mexican hat-like potential, and the third term represents the anisotropy, and gives rise to three minima on the rim of the Mexican hat, seen in Fig.~\ref{fig:barrier-specs}. The parameters were computed using Density Functional Theory in GGA+U approximation as implemented in Quantum ESPRESSO code\cite{QE-2009,QE-2017,QE20}. Brillouin zone integrations were performed on a $k$-mesh of dimensions 4$\times$4$\times$3. The scalar-relativistic PAW pseudopotentials were used within the PBE GGA exchange-correlation functional. The Hubbard repulsion on Co 3$d$ orbitals was included within the simplified DFT+$U$ formalism using a value of $U=4$~eV. The reported energies were calculated for Co spins in a ferromagnetic state, and we verified that the magnetic order with ferromagnetic Co layers aligned antiferromagnetically between each other has not led to a noticeable change in the calculated potential energy surface. This suggests a weak coupling of the magnetic order to the trimerization mode, as mentioned earlier. 

The obtained parameters are given in Table~\ref{tab:param}.
\begin{table}[t]
    \centering
    \begin{tabular}{|c|c|c|}
    \hline
        parameter & value & unit\\
        \hline
        $a$ & $-1.17$ & eV/\AA$^2$\\
        $b$ & $0.59$ & eV/\AA$^4$\\
        $\gamma$ & $-0.26$ & eV/\AA$^3$\\
        $c$ & 0.1 & eV\AA$^2$\\
        \hline
    \end{tabular}
    \caption{Model parameters for the energy surface $F(\bm{Q})$ of the trimerization mode, calculated using DFT+$U$ and reported at per formula unit value.}
    \label{tab:param}
\end{table}
The resulting free energy surface is shown in Fig.~\ref{fig:barrier-specs}(a). Zero trimerization (marked with 0) corresponds to the top of the Mexican hat, with the energy ~0.45~eV/f.u. above the minimum, shown in panel (c). The three minima are separated by the saddle points 0.25~eV/f.u. above the minima. The green line corresponds to the order parameter trajectory, corresponding to the domain wall between adjacent domains $D_1$ and $D_2$. The cuts of the free energy surface along $0-D_1$ and $D_1-D_2$ lines, obtained using nudged elastic band method with 31 images~\cite{jonsson1998nudged}, are shown in Fig.~\ref{fig:barrier-specs}(c). The spatial profiles of the trimerization amplitude $Q$ and angle $\phi$ across the wall are plotted in Fig.~\ref{fig:barrier-specs}(b). These domain walls are schematically illustrated in Fig.~\ref{fig:walls}. To obtain these profiles, we have estimated the wall width $\lambda=8.9-9.6$~nm by fitting the intensity profile across the wall from Fig.~5(g) of Ref.~\onlinecite{Xu2022multiple} and chose the order parameter stiffness $c=0.1$~eV\AA$^2$/f.u. in the free energy contribution $c(\nabla \bm{Q})^2$ to match the wall width. We then performed phase field simulations from a random initial state, which led to domain coarsening and a domain pattern with triangular vortices, where three domain walls meet at the core and the order parameter $\bm Q$ rotates by 360$^\circ$ around the vortex core, as seen in Fig.~\ref{fig:barrier-specs}(d).

\section{Conclusions}
We studied the complex interplay of ferroelectric polarization, chirality, trimerization and antiferromagnetic order in multiferroic BaCoSiO$_4$. The components of the trimerization mode, that are odd and even with respect to the inversion and mirror symmetry are improper and are governed by ferroelectric polarization and chirality. The Co/Si ionic ordering, giving rise to polarization domains, and tetrahedral rotations, inducing structural chirality, are very large and are not expected to be switchable due to large energy barriers for swapping Co and Si ions. The free energy landscape has three minima, giving rise to vortices, at which three domain walls meet and the trimerization rotates by 360$^\circ$ around the vortex core. The trimerization sets an easy direction for the antiferromagnetic order parameter (magnetic toroidal moment), which follows the trimerization rotation, although on a much longer length scale. Shear strains are not induced by the trimerization. We hope the results to stimulate further experiments on this interesting family of multiferroics. 

Acknowledgements: We acknowledge the CINECA award under the ISCRA initiative for the availability of high-performance computing resources and support.


%

\setcounter{figure}{0}
\makeatletter 
\renewcommand{\thefigure}{S\@arabic\c@figure}
\makeatother

\clearpage
\onecolumngrid
\section{Supplementary information}

\begin{figure*}[h]
   \centering
   \includegraphics[width=\linewidth]{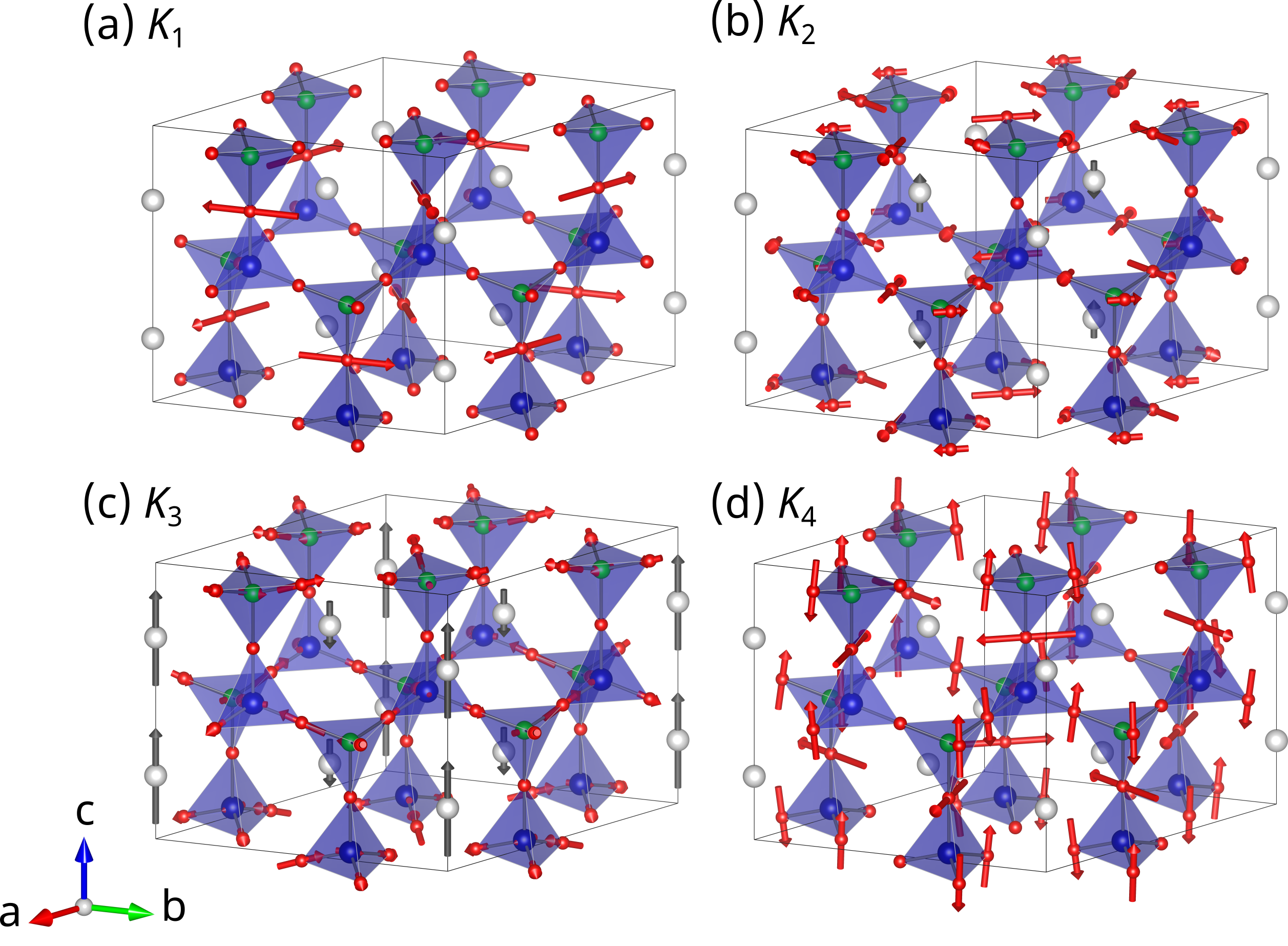}
   \caption{Zone-boundary modes at the $K$ point, $K_1$, $K_2$, $K_3$, $K_4$, present in a low-temperature trimerized structure. Ionic displacements are indicated with arrows.}
   \label{fig:k_combo}
\end{figure*}

\begin{figure}
    \centering
    \includegraphics[width=0.98\linewidth]{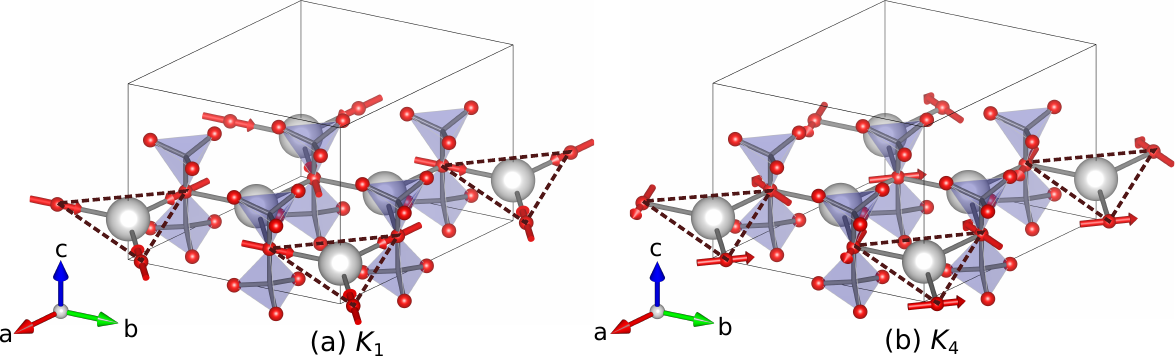}
    \caption{Displacement vector of the apical oxygen ions under the $K_1$ distortion mode.}
    \label{fig:k1_k2}
\end{figure}
\end{document}